\documentclass[journal, peerreview, onecolumn, draftcls]{IEEEtran}
\usepackage{cite}
\usepackage{color}

\ifCLASSINFOpdf
  \usepackage[pdftex]{graphicx}
\else
   \usepackage[dvips]{graphicx}
\fi
\usepackage[cmex10]{amsmath}
\begin{document}
%
\title{Interference and Congestion Aware Gradient Broadcasting Routing for Wireless Sensor Networks}

\author{\IEEEauthorblockN{Katia Jaffr\`es-Runser\IEEEauthorrefmark{1}\IEEEauthorrefmark{2},
Cristina Comaniciu\IEEEauthorrefmark{1}and
Jean-Marie Gorce\IEEEauthorrefmark{2}\\}
\IEEEauthorblockA{\IEEEauthorrefmark{1}Dept. of Electrical and Computer Engineering,
Stevens Institute of Technology, \\Hoboken, New-Jersey 07030, USA
\\ Email: {katia.runser, ccomanic}@stevens.edu}\\
\IEEEauthorblockA{\IEEEauthorrefmark{2}
Universit\'e de Lyon, INRIA \\
INSA-Lyon, CITI, F-69621, France\\
Email: {jean-marie.gorce}@insa-lyon.fr}
}

\maketitle

\begin{abstract}

This paper addresses the problem of reliable transmission of data through a sensor network. We focus on networks rapidly deployed in harsh environments. For these networks, important design requirements are fast data transmission and rapid network setup, as well as minimized energy consumption for increased network lifetime.
We propose a novel broadcasting solution that accounts for the interference impact and the congestion level of the channel, in order to improve robustness, energy consumption and delay performance, compared to a benchmark routing protocol, the GRAB algorithm. Three solutions are proposed: P-GRAB, a probabilistic routing algorithm for interference mitigation, U-GRAB, a utility-based algorithm that adjusts to real-time congestion and UP-GRAB, a combination of P-GRAB and U-GRAB. It is shown that P-GRAB provides the best performance for geometry-aware networks while the U-GRAB approach is the best option for unreliable and unstable networks.
\end{abstract}

%
\IEEEpeerreviewmaketitle

\section{Introduction}

The reliable transmission of sensed data across large-scale wireless sensor networks (WSN) has triggered lots of efforts in current research projects. Recent technologies offer low-cost and low-power chips that can be deployed for monitoring purposes in open fields. When a node senses some change in the environment, it advertises its data to one or several sink nodes. Due to the large scale of such networks, a multi-hop transmission is often used between the data source node and the sink. The routing algorithms must provide robust end-to-end transmissions, which is even more important when nodes are deployed in environments with hazardous operational conditions (e.g. high temperature, fire, humidity...). In such conditions, wireless transmissions become less reliable because of an increased number of node failures. Due to environmental effects, the transmission channel may become much more difficult, impacting the interference distribution and the congestion in the network. We show in this paper that gains in transmission efficiency are obtained upon accounting properly for either the interference distribution or the congestion in the routing decision of a gradient forwarding protocol, depending on the variability of the environment.

In this work, we target applications where the frequency of reception failures at the nodes due to external working conditions (i.e. packets losses due to channel variability or node failures) is relatively important. In this context, traditional single path approaches such as Directed Diffusion \cite{intanagonwiwat00} or Rumor routing \cite{braginsky02} are not suitable. In this case, the source-sink path easily breaks, which triggers a new flooding stage for route discovery. Route repair techniques may in that case be applied \cite{Tian03}, but such a strategy also introduces an additional delay for path repair and relies on the introduction of a specific signaling overhead.
Moreover, we concentrate on scenarios where nodes have to be deployed rapidly, as for instance for disaster relief applications and in such conditions, the network has to be up and running as quickly as possible. Hence, there is no time for complex beforehand route computations and/or rate allocation as targeted in other robust routing techniques for WSN \cite{Zhu08, Srinivasan03}.
To summarize, our goal in this paper is to propose a routing solution which adapts to the changing environment in terms of congestion/interference without triggering too much overhead at the network layer (and hence reducing the energy consumption of the network) while providing a quick network response. Reducing energy consumption and delay will affect the transmission robustness. Our design goal is to achieve the best possible robustness/energy/delay trade-off in the design of the routing strategy.

Robustness can be achieved through a hop-by-hop or an end-to-end acknowledgement procedure at the medium access or at the transport layer \cite{Stann03}. In this case, and for networks suffering from high packet losses, transmission delays due to retransmissions are increased drastically \cite{RRZhang}. Since we have a stringent delay constraint, we disregard such options which are more suited to WSNs deployed to monitor with a high fidelity a phenomenon where robustness transmission has to be guaranteed.

Another mean of increasing the robustness is by introducing totally redundant transmissions, either by defining a-priori fixed redundant source to sink paths in the network or by allowing spontaneously one or more nodes at a time to forward a packet given a set of forwarding policies. In the first case, the forwarding decision is set once and for all by the routing protocol and in the second case, depending on the state of the network, nodes decide locally to forward or not their packet.

Redundant source-sink paths are constructed in braided multi-path routing algorithms \cite{ganesan01, bush05}, which are multi-path versions of Directed Diffusion. In these works, N routes are reinforced after the flooding stage and maintained with either `keep alive' packets \cite{ganesan01} or by alternatively sending the data in a round robin manner on each path to reduce the route maintenance load \cite{bush05}. Such solutions, where routes have to be directly defined also introduce additional delay in the set up stage of the network.

To provide a rapid sensor network roll-out and quick transmissions, gradient broadcasting techniques are the most promising solutions (cf. \cite{Maroti04,GRAB05,Chen05,Lim05, Poor}).
In these approaches, no routes are set prior to sending data, and only costs are assigned to nodes being equal to the minimum cumulative link cost to the sink node.
All the costs create a discrete gradient field whose minimum is located at the sink (similar gradient fields are also exploited in a different manner to obtain efficient anycast routes in mobile networks \cite{lenders2006}). When a sensor has some data to send, it broadcasts its data packet by assigning its own cost to the packet cost $Q_p$. The neighbor nodes {\it with costs smaller than the packet cost $Q_p$} decide to broadcast the packet or not, based on a set of forwarding rules. If the packet is broadcasted, its cost $Q_p$ is updated with the cost of the forwarding node. All subsequent transmissions always `roll down the hill' to reach the sink. When several sink nodes exist, several cost fields are determined. The cost field is either set up by an a-priori flooding stage \cite{Maroti04,GRAB01,GRAB05,Lim05} or on-demand with a request/response packet exchange \cite{Poor,Chen05}.

The network setup is fast since a single flooding stage is needed to create the cost field as shown in \cite{GRAB01}. No complicated procedure is needed to create routes from sources to sink and hence no additional route repair procedures are implemented to guarantee robustness. Also, to deliver a fast network response, no hop-by-hop or end-to-end acknowledgements are considered. The statistical redundancy obtained to multiple packet forwards is supposed to provide enough robustness for the information to be gathered at the sink. Of course, there is a price to pay in terms of energy since more nodes participate in the forwarding effort.
Hence, the policy chosen to decide whether a node closer to the sink forwards or not a message impacts the robustness/latency/energy trade-off and constitutes the heart of the protocol.
Current gradient broadcasting algorithms do not consider the impact of interference or congestion in the policy. For instance, the state of the art solution known as GRAB \cite{GRAB05} proposes a policy that creates a forwarding mesh whose structure depends only on the relative distance of the forwarding node to the sink node.

In this work, we propose three different broadcasting policies which account for interference and/or congestion: P-GRAB, U-GRAB and UP-GRAB, denoting `Probabilistic-GRAB', `Utility-GRAB' and `Utility and Probabilistic GRAB, respectively.
In these algorithms, the broadcasting decision is taken considering additional information to improve the robustness/latency/energy tradeoff. This additional information is either retrieved by inducing cooperation among the nodes (U-GRAB and UP-GRAB) or by making them act independently (P-GRAB). In P-GRAB, nodes do not interact and use their knowledge of the topology of the network to estimate their interference impact on the network. In the two other policies (U-GRAB and UP-GRAB), nodes interact following a utility-based model to adjust their broadcasting decisions to the congestion of the channel.

P-GRAB accounts for the interference impact of the broadcasting decision on its neighbor vicinity. The aim of this broadcasting scheme is to favor the re-transmission for nodes that belong to a relatively less dense region of the network. The algorithm assigns a probability of forwarding a packet to a sensor by accounting for the impact of the broadcasting decision on the interference distribution of the network \cite{iswpc08}.
To better adjust to the dynamics of the network, the U-GRAB policy accounts for an estimate of the current channel congestion by defining a heuristic utility-based algorithm where nodes decide to broadcast a packet or not, depending on their energy level and the channel occupancy. As we have mentioned before, cooperation is also encouraged via a pricing function included in the utility definition. The congestion oriented strategy is implemented in the `U-GRAB' algorithm \cite{milcom09}.
The third policy, UP-GRAB, introduces the interference impact metric defined in P-GRAB into the U-GRAB algorithm to account for both the channel congestion and the interference impact of the broadcasting decision.

These approaches are tested in the following regarding the severeness of the working conditions by considering a probability of failure as proposed in \cite{GRAB05}. We define a failure as being the event that a message can not be relayed by a sensor. This originates from two different cases: $i)$ the sensor can not not receive the packet because of fading or $ii)$ there is a hardware failure of the sensor that can not receive or send a packet.
Transmission errors due to interference or congestion which depend on the geometry of the network and on the traffic characteristics are accounted for using the discrete time event OMNet++ simulator in a modified version of the SENSIM simulator \cite{LSUsimulator}. To model the additional outages due to a harsh environment, we model the outages due to fading and sensor node failure with a unique probability of failure referred to as $p_f$. Hence, when a message is received by a sensor knowing the interference and congestion status of the network, there is an additional probability $p_f$ that the message is not received and hence not forwarded by the node. More complex failure models may be implemented in the future, but the simple one considered here already provides a good insight on the targeted protocol's performance.

This paper is structured as follows: Section \ref{sec:gradientbroadcasting} outlines related work on gradient broadcasting and gives a short description of the GRAB algorithm. The first probabilistic approach P-GRAB is presented in section \ref{sec:PGRAB} and the utility-based approach U-GRAB in section \ref{sec:GGRAB}. The hybrid version UP-GRAB is detailed in section \ref{sec:PGGRAB}. Extensive simulation results comparing the three algorithms are provided in section \ref{sec:results}. The last part concludes the paper.

\section{Gradient broadcasting}\label{sec:gradientbroadcasting}

\subsection{Background}\label{subsec:background}

The basic version of a gradient broadcasting algorithm, referred to as BGB in this work, is described in \cite{Maroti04} and \cite{Poor}. There is no additional policy and as long as a node is closer to the sink (i.e. has a smaller cost value than the packet cost), it forwards the packet. This approach triggers a significant number of broadcasted packets but in turn it delivers the packets quickly. Robustness is also high due to the high number of packets sent even if a significant number of collisions occur. Of course, a short term memory stores the identification of the previously transmitted messages to avoid re-transmitting twice the same packet.

Mar\'oti provides in \cite{Maroti04} a global framework for such broadcasting algorithms referred to as `Directed Flood-Routing Framework' where the decision policy of a node is modeled as a finite state machine. In addition to presenting the basic version, it proposes a `Fat Spanning Tree' implementation where the finite state machine provides rules enabling the 1-hop neighbors of the nodes that belong to the shortest path from source to sink nodes to broadcast the packet. Such approach is more robust to failure than a regular shortest single-path routing protocol. However, he considers the spanning tree to be constructed in an a-priori flooding stage which adds another expensive flooding step to the protocol.

Chen et al. \cite{Chen05} use the cost field in a different manner as they do not impose strict rules but a selection process of the node that forwards the packet. Their protocol, called Self-Selective Routing (SSR) uses a back-off delay and two broadcast packets to elect the node that forwards the data packet. All the nodes receiving a packet to forward from a node $A$ start a back-off timer proportional to their cost. The closest node to the sink, node $B$, is the first to send its packet as its backoff is the smallest one. This packet is understood by the neighboring nodes of $B$ as an implicit acknowledgement. To warn the neighbors of $A$ that don't get the implicit acknowledgement from $B$, node $A$ sends an explicit ACK packet upon reception of the data packet sent by $B$. This process is robust as paths adjust to failures. It also reduces drastically the number of forwarded packets. But the self-selection process increases the end-to-end delay and the radio layer must listen to each contention phase to get the appropriate information which increases energy expenditure.

\subsection{The GRAB algorithm}

The GRAB algorithm has been proposed by Ye et al. \cite{GRAB01, GRAB05} to provide a robust routing algorithm based on the gradient broadcasting concept. The GRAB algorithm is a solution for reducing the number of forwarding nodes by assigning a credit to each packet transmission in the network and setting rules on the credit consumption at each hop.
The GRAB algorithm \cite{GRAB05} starts by a flooding stage where each node gets a cost proportional to a distance measure to the sink. Flooding is started by the sink sending an advertisement (ADV) packet containing its own cost ($Q=0$ for instance). All the other nodes have an initial cost $Q=+\infty$. A node $A$ with cost $Q_A$ that receives an ADV packet with packet cost $Q_p$ updates its own cost if $Q_p + L < Q_A$, $L$ being the link cost. If this condition is met, the new cost $Q_A$ is set to $Q_p + L $ and a new ADV packet is sent with a new packet cost $Q_p = Q_A$. To reduce the flooding load, a back-off timer proportional to $Q_A$ is decremented before sending the ADV packet. Consequently, the node with the lowest value of $Q_A$ sends its packet first, acting as an implicit acknowledgement that prevents other nodes with higher costs from forwarding their ADV packet. With this algorithm, only one ADV packet per node is sent in the cost field setup stage (cf. \cite{GRAB01} for the proof).

The link cost value can be expressed in various metrics (in hops, in meters, etc)... In this work, we propose an energy related link cost which is proportional to the power needed to transmit a packet to the neighboring node the ADV packet is coming from. The resulting power is the opposite of the pathloss expressed in decibels between both nodes. Hence, the link cost between node $i$ and $j$ separated by $d_{i,j}$ meters is given by:
\begin{equation}\label{eq:linkCost}
    L(i,j) = 10\cdot \alpha \cdot \log_{10}( d_{i,j})
\end{equation}

If the nodes know their positions, a node $j$ receiving an ADV packet from node $i$ can calculate $d_{i,j}$ if node $i$ includes its location in the ADV packet.
When the location information is not available, which is often the case for rapidly deployed networks, this energy metric can still be computed by the nodes and with the same overhead. In that case, node $i$ explicitly includes the value of its transmission power in the ADV packet. Upon receiving this packet, node $j$ can easily determine the value of the pathloss using:
\begin{equation}\label{eq:linkCost2}
    L(i,j) = \left( P_t^i\right)_{dB} - \left(P_r^j\right)_{dB}
\end{equation}
\noindent where $\left( P_t^i\right)_{dB}$ is the transmission power of node $i$ and $\left(P_r^j\right)_{dB}$ the reception power at node $j$, expressed in dB, measured by the physical layer of the protocol stack.

\begin{figure}
    \centering
  \includegraphics[width=3.5in]{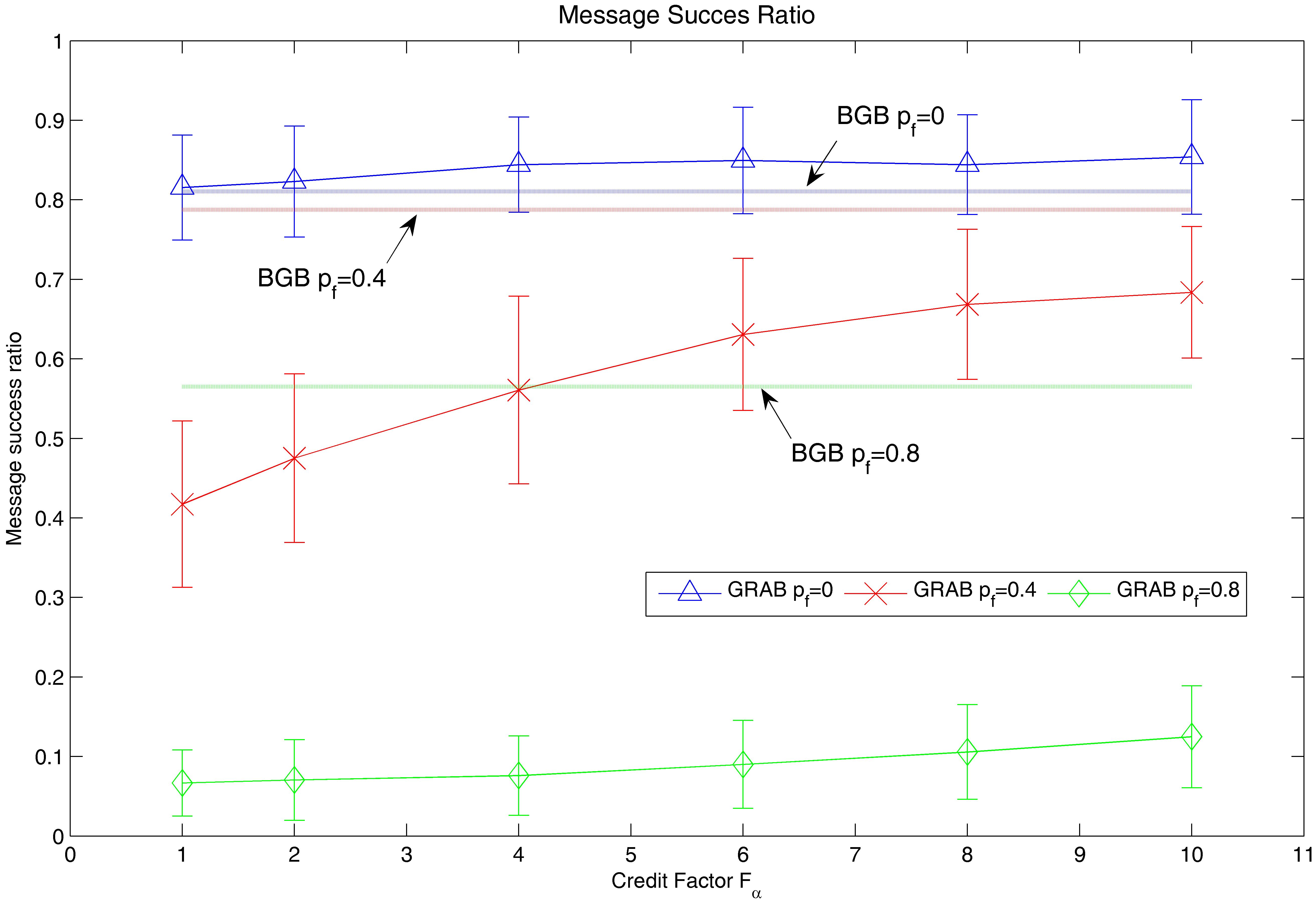}
    \caption{Comparison of BGB and GRAB regarding robustness for several values of the Credit Factor $F_\alpha$ and a probability of node failure $p_f\in[0,0.4,0.8]$. The credit factor $F_\alpha$ is here applied to the cost of the source node $Q_S$ to calculate the initial credit value $\alpha$ with $\alpha = F_\alpha \cdot Q_S$.}
    \label{fig:Grab-BGrab-Rob}
\end{figure}

\begin{figure}
    \centering
    \includegraphics[width=3.5in]{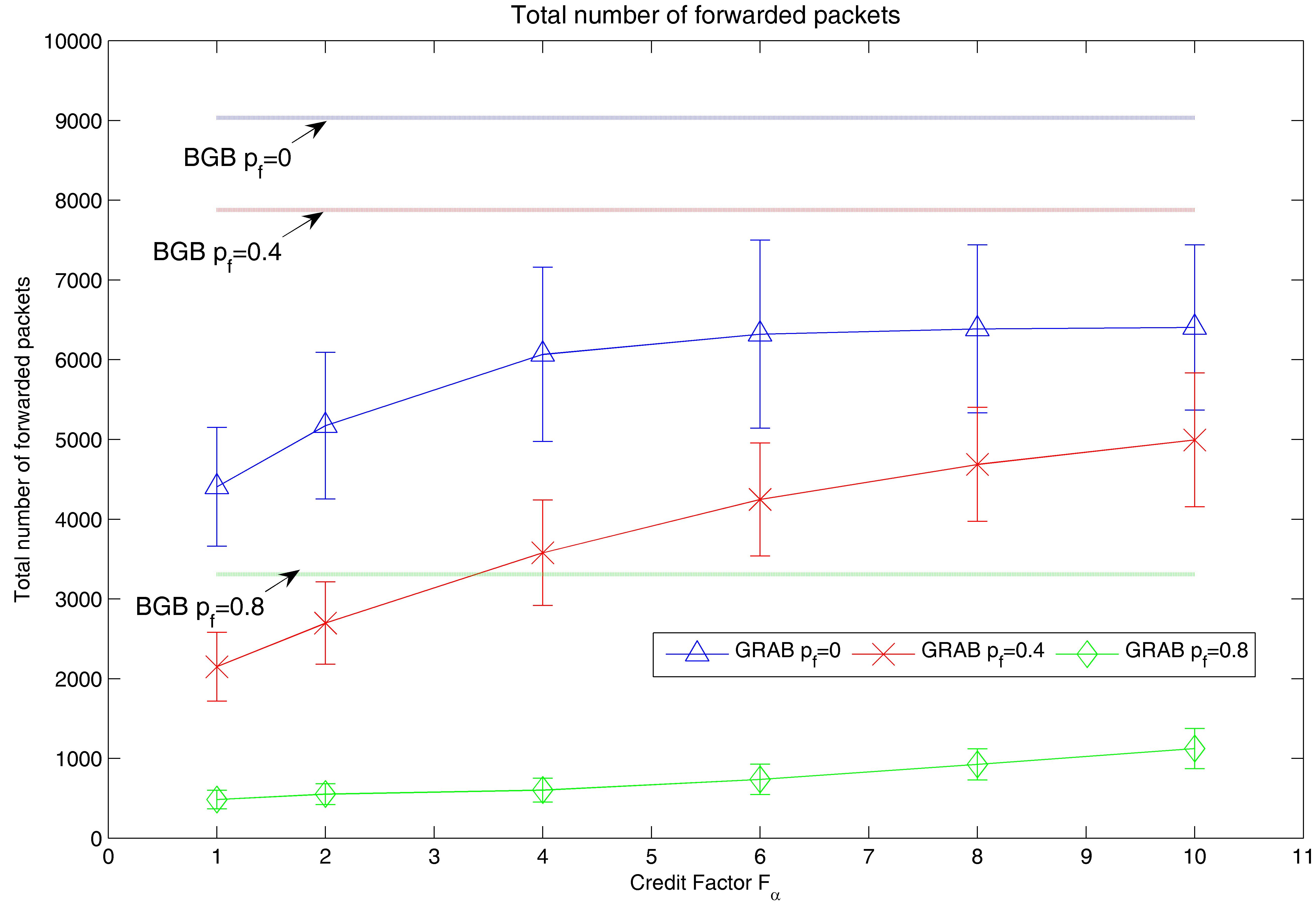}
    \caption{Comparison of BGB and GRAB regarding the total number of forwarded messages (right) for several values of the Credit Factor $F_\alpha$ and a probability of node failure $p_f\in[0,0.4,0.8]$. The credit factor $F_\alpha$ is here applied to the cost of the source node $Q_S$ to calculate the initial credit value $\alpha$ with $\alpha = F_\alpha \cdot Q_S$.}
    \label{fig:Grab-BGrab-For}
\end{figure}

To reduce the overhead of multiple forwards, the goal of GRAB is to provide a forwarding mesh with fewer nodes forwarding near the source and the destination and more relays being active in between. Therefore, GRAB assigns a credit to each data packet that is sent out by a source node. Each forwarding node determines if it has enough credit to broadcast the packet relatively to the distance to the destination node. If there is enough credit left, the nodes broadcast the packet to reach a fixed number $N_n$ of nearest neighbors through power adjustment. On the contrary, if credit is low, it only forwards to its nearest neighbor. This creates a forwarding mesh whose width can be adapted by the choice of the credit or of the number of nearer neighbors. We implemented this algorithm as a benchmark for our proposed protocols.

Performance of GRAB is being compared to the performance of BGB in Fig.~\ref{fig:Grab-BGrab-Rob} and \ref{fig:Grab-BGrab-For}. The credit factor $F_\alpha$ represents the additional credit the source adds to the first packet sent in GRAB. Hence, $F_\alpha$ is here applied to the cost of the source node $Q_S$ to calculate the initial credit value $\alpha$ according to $\alpha = F_\alpha \cdot Q_S$.
As presented in Fig.~\ref{fig:Grab-BGrab-Rob} and \ref{fig:Grab-BGrab-For}, the performance simulations of GRAB and BGB show that GRAB needs fewer transmitted packets compared to BGB for reliable environments, i.e when the probability of node failure is low. However, when the probability of node failure increases, the amount of initial credit of GRAB has to be increased to improve the robustness of the transmission. In this case, GRAB tends to use as much redundancy as BGB, which increases drastically the number of forwarded packets, and thereby the energy expenditure of the network.
GRAB is also about 2.5 times slower than BGB in delay which is due to the power adjustment feature of GRAB that only broadcasts the packets to reach a fixed number of $N_n=3$ neighbors.
To overcome the limitations for the GRAB algorithm, our aim in this work is to propose a new gradient based routing strategy that:
\begin{itemize}
\item Reduces the number of forwarding nodes when the network is reliable,
\item Increases the transmission robustness when the network becomes unreliable and
\item Provides faster sensor-sink communications.
\end{itemize}

\section{The P-GRAB algorithm}\label{sec:PGRAB}

The proposed P-GRAB gradient broadcasting heuristic aims at accounting for both energy and interference within its forwarding policy. Hence, the forwarding decision of a node is made knowing the potential interference it creates when broadcasting a packets and the energy remaining in its battery. The rationale is to favor nodes creating low interference to transmit for a network where all the nodes have the same level of energy in their batteries. As energy depletes, nodes located in relatively sparser regions stop transmitting because of their energy constraint and nodes in relatively denser regions with higher energy can broadcast their packets.

In the rest of the paper, we consider a network of $N$ sensors where the costs are already set according to the cost setup algorithm of GRAB using the cost metric of Eq.~\ref{eq:linkCost} or Eq.~\ref{eq:linkCost2}. The broadcasting decision follows a probability of forwarding a packet $P_{FW}$. When a node $n_i$, based on its cost value $Q_i$, is allowed to forward a packet, it forwards the packet with probability $P_{FW}$ defined as follows:
\begin{equation}\label{eq:pFwd}
    P_{FW} = P_{IA} * P_{LD}
\end{equation}
\noindent where $P_{IA}$ is a measure of the probability of interference avoidance and $P_{LD}$ is the probability of life duration of the node.

\subsection{The probability of Life Duration}

$P_{LD}$ is a function of the remaining energy in the node and the way the energy has been spent in the past. Each node can easily store the number $N_F$ of already broadcasted messages, and its initial energy $\mathcal{E}_{initial}$. It also knows the remaining energy $\mathcal{E}_{remaining}$ in its battery. Before forwarding a packet, the node can estimate how many transmissions $N_{EF}$ it can still do with:
\begin{equation}
N_{EF} = \mathcal{E}_{remaining}/\mathcal{E}_{F}
\end{equation}
\noindent where $\mathcal{E}_F$ is the energy spent per packet broadcast $\mathcal{E}_F = (\mathcal{E}_{initial} - \mathcal{E}_{remaining})/N_F$.

The probability of life duration $P_{LD}$ is defined using:
\begin{equation}\label{eq:PLD2}
P_{LD} = 1 - 1/(N_{EF}+1)
\end{equation}
$P_{LD}$ is close to one when the number of expected forwards $N_{EF}$ is high and tends to zero when no packets can be broadcasted anymore ($N_{EF}=0$).

This model is really simple and does not account for the leakage of the battery which is not a linear function of time and the remaining energy. It means that the node is overestimating the value of $N_{EF}$. But since this value is updated at each new packet reception, the discrepancy between the real value of $N_{EF}$ and the estimated one is kept small.

\subsection{The probability of Interference Avoidance}

This probability favors the broadcasting for nodes whose impact in terms of interference on its vicinity is less important relatively to the interference impact of its 1-hop neighbors.
The denser the vicinity of a node is, the more nodes its transmission reaches and can potentially disturb. We define here a relative measure of neighborhood density that considers the difference in size of a node's neighborhood relatively to the size of the neighborhoods of its 1-hop neighbor nodes.
Let $N_i$ be the number of neighbors of node $i$. A node $v_j$ is a neighbor of $i$ if the received power $\mathcal{P}_{ji}$ at location $j$ from node $i$ verifies $\mathcal{P}_{ji} > \mathcal{P}_{lim}$ with $\mathcal{P}_{lim}$ the sensitivity of a sensor.
A node $i$ can estimate the average discrepancy between its neighborhood size $N_i$ and the neighborhood size $N_j$ of each 1-hop neighbor $j$ for $j\in [1..N_i]$. This measures how many more (or how fewer) sensors the node $i$ may interfere on average than its neighbor nodes.
This average neighborhood discrepancy $\Delta(i)$ of node $i$ is defined by:
\begin{equation}\label{eq:delta}
\Delta(i) = \frac{\sum_{j=1}^{N_i} (N_i-N_j)}{N_i}
\end{equation}

If $\Delta(i)=0$, the node distribution is uniform. If $\Delta(i) > 0$ (resp. $\Delta(i) < 0$), node $i$ belongs to a denser (resp. sparser) neighborhood than its 1-hop neighbors. That does not mean that node $i$ belongs to a sparse region, but compared to its 1-hop neighbors, its vicinity is less dense than theirs.

\begin{figure}
\centering
  \includegraphics[width=3.5in]{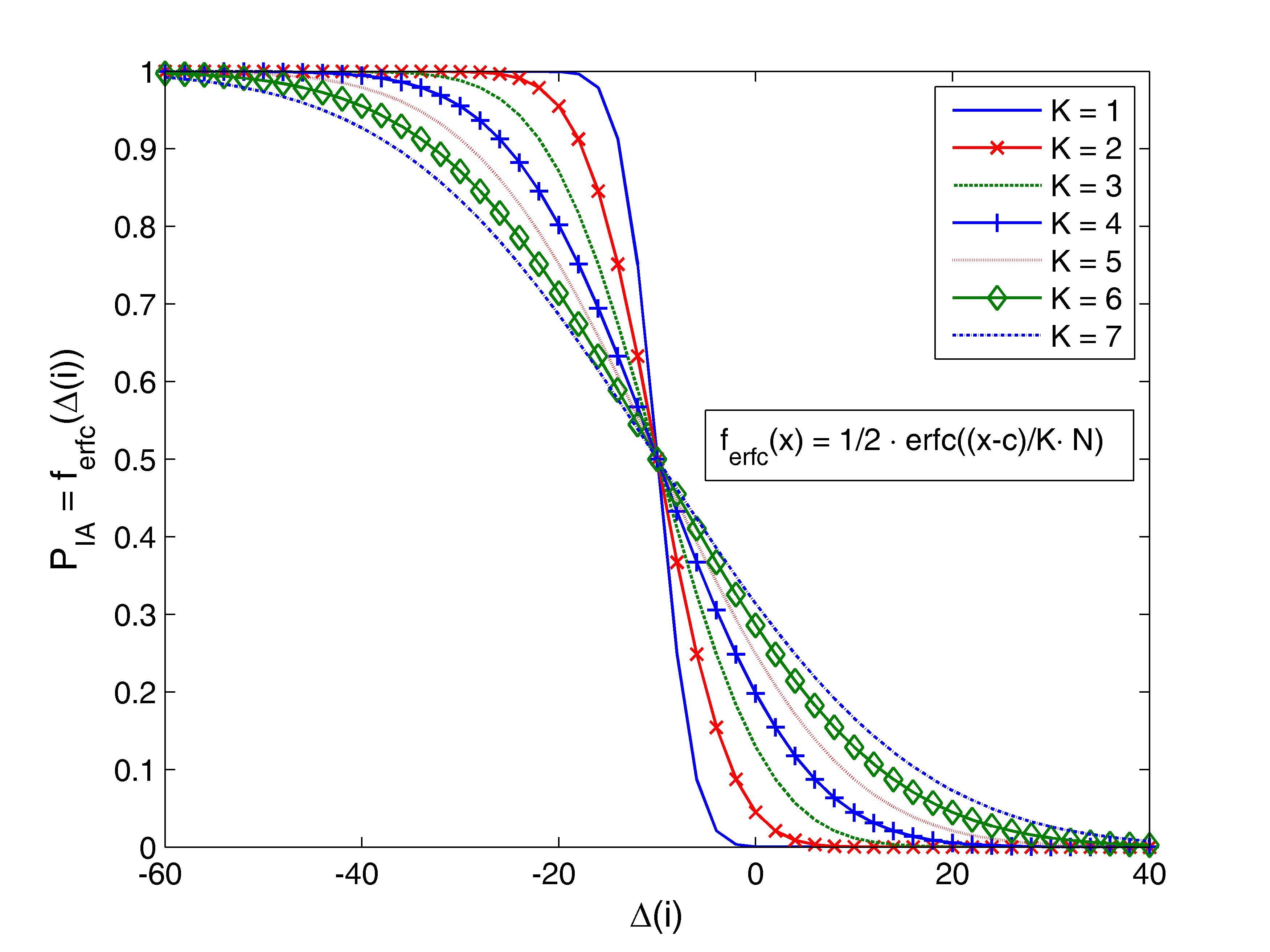}
    \caption{Conversion function $f_{\rm erfc}$ for $\Delta_{min}=-60$, $\Delta_{max}=40$, $c=-10$ and various spreading factor values $K$. We have $N=100/24$ since we select a target interval of [a,b]=[-12,12] for the standard erfc(x) function.}
    \label{fig:fIAErfc}
\end{figure}

To favor the forwarding of nodes that provide less interference to their neighborhood, we define a conversion function denoted $f_{\rm erfc}$ that assigns a probability $P_{IA}$ to any possible value of $\Delta(i)$. This function relies on the complementary error function $x \rightarrow {\rm erfc}(x)=\frac{2}{\sqrt{\pi}} \int_x^\infty e^{-t^2}dt$ that is scaled to fit to the set of $\Delta(i)$ values. $f_{\rm erfc}$ is shown in Figure \ref{fig:fIAErfc} and defined by:
\begin{equation}\label{eq:fIAErfc}
    f_{ \rm erfc}(\Delta(i)) = 1/2 \cdot {\rm erfc}[(\Delta(i)-c)/m]
\end{equation}

with $c$ the center value of the interval $[\Delta_{min}, \Delta_{max}]$, with $\Delta_{min}$ and $\Delta_{max}$ the minimum and maximum of the $\Delta(i), i\in[0..N]$ values, respectively. We also set $m=K \cdot n$ with $n = (\Delta_{max}-\Delta_{min})/(b-a)$ a scaling value that transforms the interval $[a,b]$ of the basic erfc function into $[\Delta_{min}, \Delta_{max}]$. In the following, we select the part $[a,b] = [-12,12]$ of erfc$(x)$ to provide the outcome of Fig.~\ref{fig:fIAErfc}. The $K$ factor, $K\geq1$, spreads or shrinks the function.

The motivation in defining this specific conversion function is to permit nodes which create low interference to have a high probability of forwarding and reciprocally, to reduce the forwarding effort for nodes located in a dense area. Also, it is possible to adjust the shape of the erfc function using the parameter $K$ (as presented in Fig.~\ref{fig:fIAErfc}) to get closer to a ramp function. For low values of $K$, the choice is almost binary and a node located in a sparser area transmits while a node in a more dense area drops the packets. For higher values of $K$, the probability better accounts for the relative value of $\Delta_(i)$ and hence better fits to the geographical distribution of the nodes.
To summarize, for small values of $K$, the nodes that yield high interference almost never forward a packet while nodes that belong to sparse neighborhoods almost always forward. For high values of $K$, each node gets a different forwarding probability varying almost linearly between zero and one.

\subsection{About the distributed implementation of the protocol}

There are several ways of determining the average neighborhood discrepancy measure of Eq.\ref{eq:delta} and the value of $P_{IA}$ of Eq.\ref{eq:fIAErfc} by the nodes. Each node needs to know the number of neighbors $N_j$, $j\in[1..N_i]$ of its 1-hop neighbors and the maximum and minimum values of $\Delta(i)$ (i.e. $\Delta_{min}$ and $\Delta_{max}$). Computation may be performed locally, in a centralized manner or distributively.

%

Here, the computation effort is distributed among the nodes. In this case, a node simply advertises an estimate of its own number of neighbors once the cost field setup stage is finished. During the cost field setup stage, a node keeps track of every neighbor ADV packet it receives by storing the neighbor node ID in a neighboring table. At the end of the cost field setup stage, each node has a good estimate of its own number of neighbors since each node advertises at least once its cost value. Then, once the cost field is created, each node $i$ can advertise its own number of neighbors $N_i$ by broadcasting it. This additional broadcasting stage relies on one packet broadcast per node and hence has the same small overhead as the cost field setup stage. This neighborhood discovery protocol relies of course on the assumption of a symmetric transmission channel. The last information needed by the distributed implementation concerns the knowledge of the $\Delta_{min}$ and $\Delta_{max}$ values. We recall that these values are needed to scale the erfc conversion function of Eq.~\eqref{eq:fIAErfc}. Minimum and maximum bounds for $\Delta(i)$ values can be computed knowing the distribution of the nodes and then can be broadcasted to the sensors by only slightly increasing the ADV packet size.

The drawback of this approach is that collisions of the ADV packets may result in an underestimation of the $N_i$ values and thereby reduce the accuracy of the $\Delta(i)$ measures. Consequently, this distributed approach provides better results for reliable networks, for whichthe estimation error for the $\Delta(i)$ is low. Our simulation results in section \ref{sec:results} show that P-GRAB has the best performance for reliable networks. In this context, it is reasonable to consider a distributed estimation for $\Delta(i)$.

\subsection{Impact of the spreading factor $K$}\label{subsec:Kvalue-PGRAB}

\begin{figure}
  \vspace{0.1in}
  \includegraphics[width=3.5in]{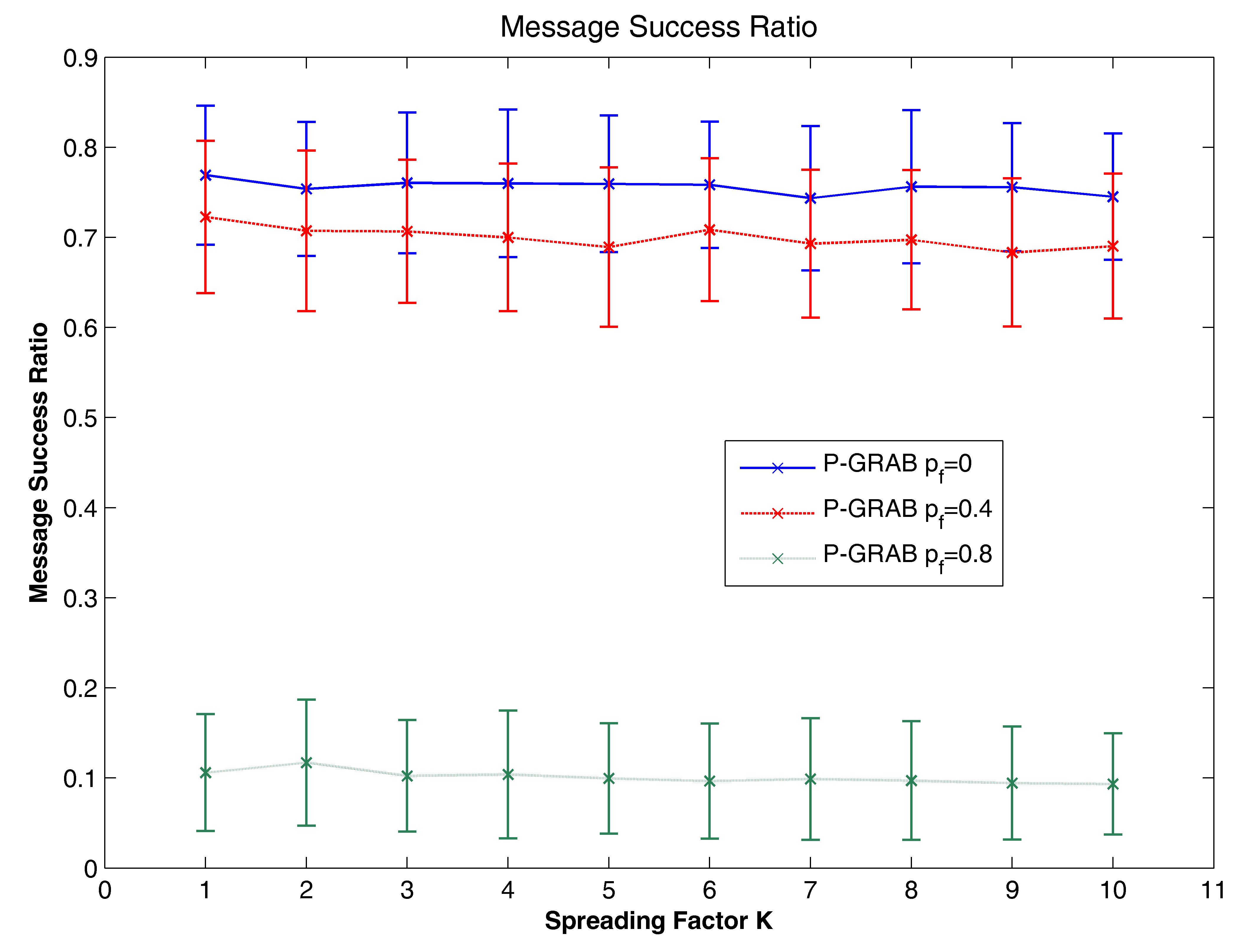}
  \vspace{0.1in}
  \includegraphics[width=3.5in]{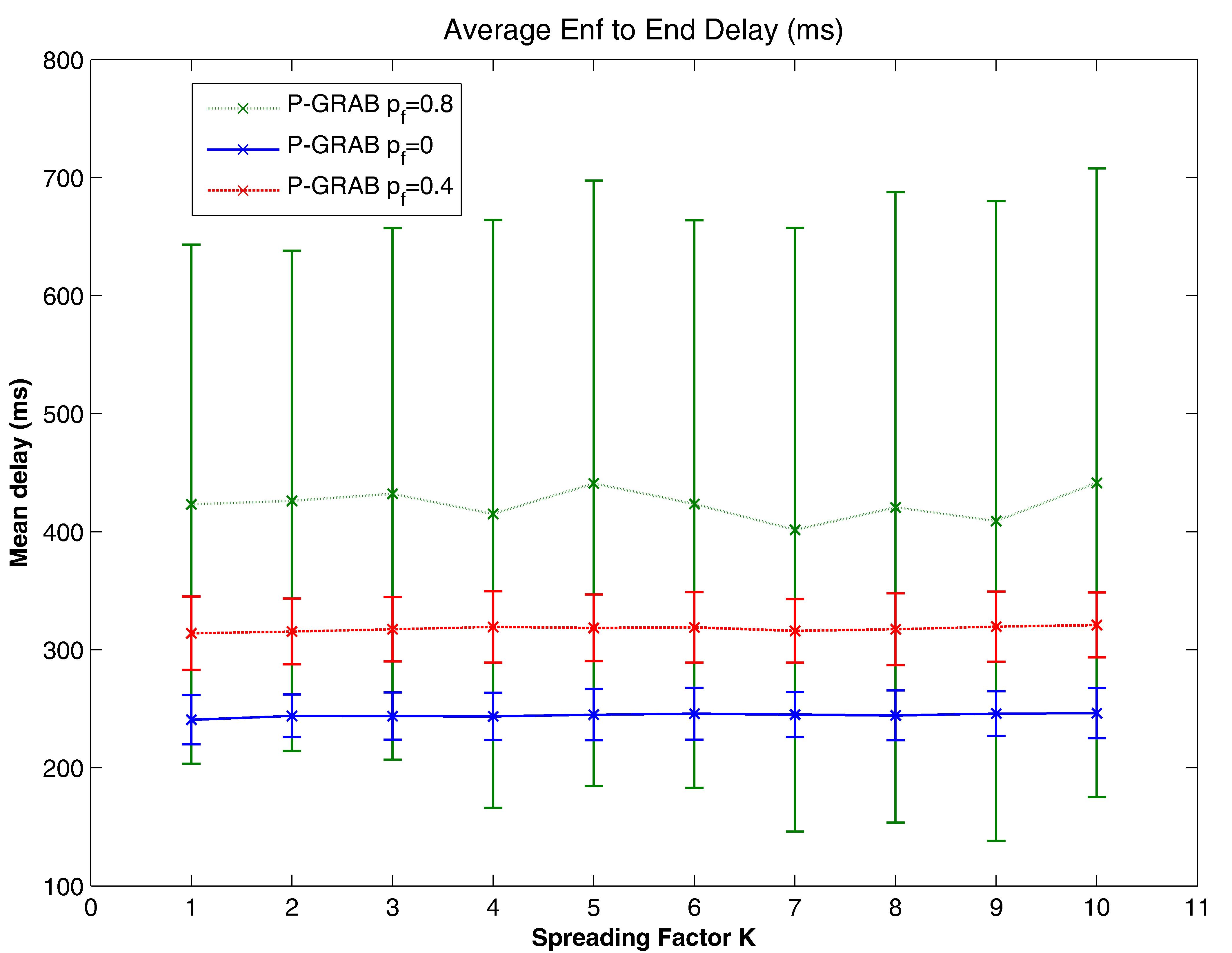}
  \vspace{0.1in}
  \includegraphics[width=3.5in]{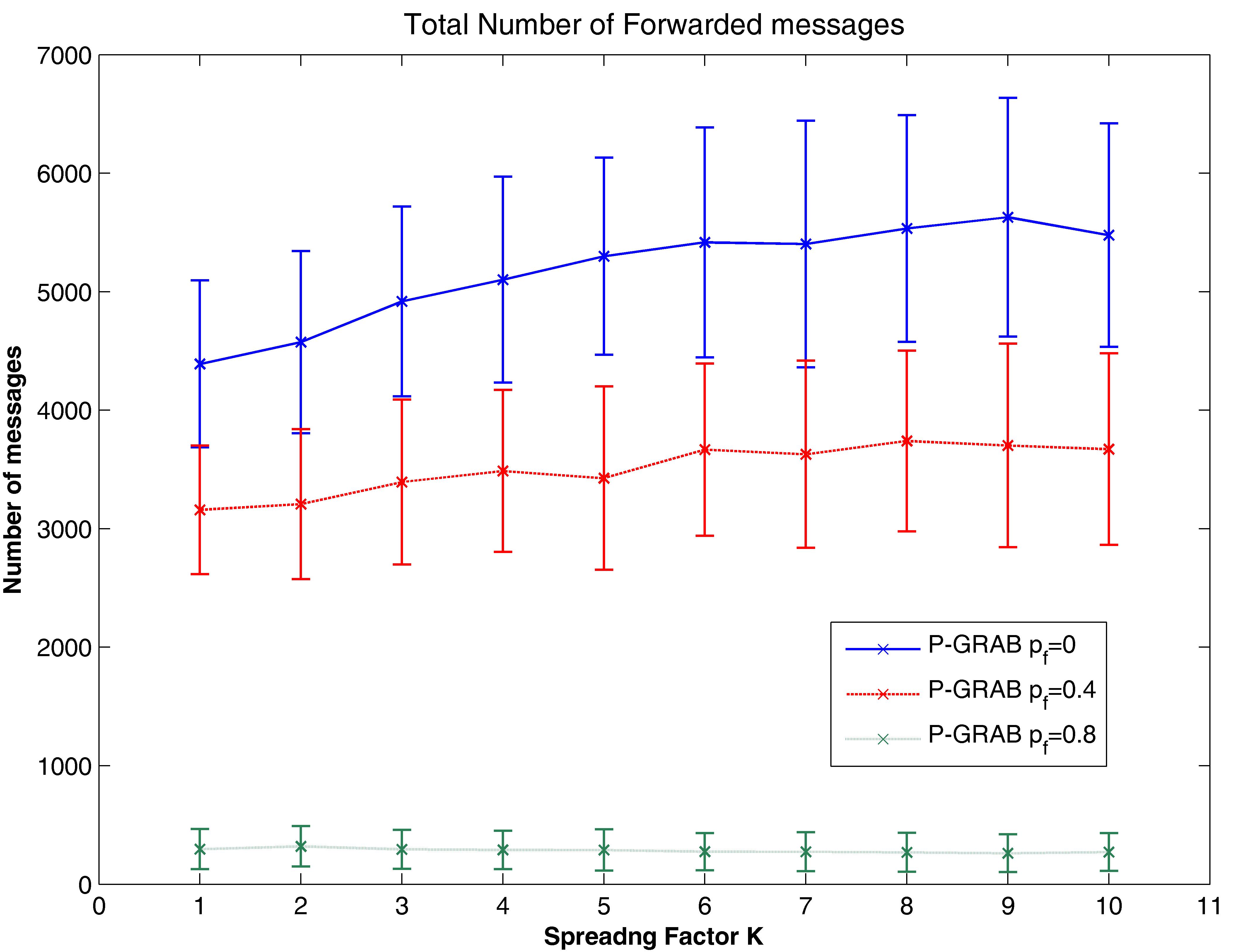}
  \caption{Impact of the spreading factor $K$ on robustness (top), delay (middle) and the number of forwarded packets (bottom).}
  \label{fig:resPGRABKParam}
\end{figure}

The impact of a unique spreading factor $K$ assigned to all the nodes has been investigated with robustness, energy and end-to-end delay metrics. The complete simulation setup is detailed in Section \ref{subsec:simSetup}. Results are presented in Fig.~\ref{fig:resPGRABKParam}.

The parameter $K$ has little influence on the robustness or the delay. A slight increase of robustness can be noticed with an increase in the $K$ value. The strongest impact is observed for the number of forwarded packets which increases with $K$. For larger values of $K$, more nodes with higher $\Delta(i)$ values will be able to forward while the nodes with lower $\Delta(i)$ values will see their forwarding probability decrease. The overall impact is that this results in an increase of the number of forwarded messages. As nodes that trigger more interference to their neighborhood forward more often for high values of $K$, more collisions occur explaining the slight increase in robustness. It is for small values of $K$ that the algorithm has the best energy/robustness trade-off as fewer packets are needed to achieve the same robustness. In this case, the conversion function is close to a simpler ramp model that can be easily implemented on a sensor node with a low computational capacity. Based on this result, we set $K=2$ for all the following instances of the $P-GRAB$ algorithm presented in this work.

As the network evolves, sensors running out of power disappear and the geometry of the network changes. Hence, as defined in the original GRAB protocol, the cost values are updated by an ADV request of the sink. During this stage, the $P_{IA}(i)$ values are computed again using the distributed approach mentioned earlier. However, this adjustment to the geometry change of the network is done at a slow pace. It also does not account for the small-scale variations due to short outages of the nodes. To provide a more reactive routing strategy, we introduce U-GRAB in section \ref{sec:GGRAB}, whose aim is to adjust to the congestion level of the network on top of accounting for the energy depletion of a node.
Assigning the same value for $K$ to each sensor in the network might not trigger the best possible outcome since it does not account for the local density of a set of neighbor nodes. The issue of determining a specific value of $K$ for each sensor is addressed by the UP-GRAB algorithm presented in section \ref{sec:PGGRAB}.

\section{The U-GRAB algorithm}\label{sec:GGRAB}

The choice of the forwarding probability in P-GRAB is mostly conditioned by the spatial distribution of the nodes. Since it does not account in real time for the data flows and the local congestion of the channel, the nodes do not adjust to the dynamics of the network. In this second proposed policy, we aim at accounting for an estimate of the current channel congestion and the energy level of the nodes using a heuristic utility-based algorithm.

The proposed utility-based distributed policy can be analyzed within a game theoretical framework. The corresponding game is a non-cooperative game where the cooperative behavior of the nodes is enforced by means of a pricing function. This function selectively rewards nodes for forwarding based on their energy level and on the congestion of their channel. In such game theoretic models, the players of the game are the wireless nodes, their actions are their choices of transmission parameters (e.g., transmission powers, channels, access probability, backoff delays or next hop nodes) and their utilities are performance measures that each terminal is trying to maximize \cite{comaniciu07}. The nodes choose their actions independently, but their choice impacts on all the users of the network. The players of the game are assumed to be rational, acting in their best interest to maximize their own utility.
Cooperation among users has been often introduced within a non-cooperative game by modifying the individual utility for the nodes using a pricing function. Such modification can enforce cooperative behavior.
The considered game is closely related to the Santa Fe Bar Problem (SFBP)\cite{mishra} where a congested resource, the bar, is shared by a set of agents, the bar customers.

We show that the exact derivation of the game equilibrium and the determination of a closed-form expression of all the parameters of the utility function is tedious. We propose instead a heuristic, utility-based distributed algorithm where the nodes adjust the parameters of their own utility functions depending on their current energy level. The gradient cost field is still set up as proposed in GRAB. It is the forwarding stage of U-GRAB that differs from GRAB: once a node has the proper cost for broadcasting a packet, it decides to forward or not based on the utilities it gets knowing the congestion of the channel and its own remaining energy.

\subsection{An utility-based forwarding policy}

We consider a network of $N$ sensors. Each sensor $n$ shares the wireless channel with a set of neighbor nodes denoted $\mathcal{V}_n = {1,\dots, V}$, with $v \in \mathcal{V}_n$. When a node $n$ receives a packet of cost $Q_P>Q_n$, the node has to choose a strategy $s_n$ in the strategy set $S =\{0, 1\}$ where:
\begin{itemize}
\item $s_n=1$ corresponds to the action {\it Forward} packet,
\item $s_n=0$ corresponds to the action {\it Drop} packet.
\end{itemize}

Let $c_{\mathcal{V}_n}$ be the level of congestion of the wireless channel used by node $n$ and its neighboring nodes in vicinity $\mathcal{V}_n$. The level of congestion for the channel is defined as the number of concurrent accesses on the channel at time $t$.
In addition, let $c$ denote the congestion limit of the channel. The channel is considered congested if $c_{\mathcal{V}_n} \geq c$ and not congested if $c_{\mathcal{V}_n} < c$. Let $0 \leq \alpha_n \leq 1$ denote the benefit value to sensor $n$ for using the channel to forward a message and $0 \leq r_n \leq 1$, the reward for not wasting energy in rebroadcasting a packet. The utility function for node $n$ is then given by:
\begin{equation}
u(s_n , c_{\mathcal{V}_n}) = \left\{
    \begin{array}{l l}
		\alpha_n 	    & {\rm if}~s_n = 1~{\rm and}~c_{\mathcal{V}_n}< c,\\
        \alpha_n - 1	& {\rm if}~s_n = 1~{\rm and}~c_{\mathcal{V}_n}\geq c,\\
  		r_n 	        & {\rm if}~s_n = 0.\\
    \end{array}
\right.
\label{eq:payoff1}
\end{equation}

Each sensor chooses the action that yields the maximum utility, knowing the values of the rewards and the congestion status of the channel. The choice of $\alpha_n$ and $r_n$ shapes the behavior of the nodes. Hence, the expected utilities depend on the particular strategic choice $s_n$ of node $n$, its reward $\alpha_n$ for using the channel, its reward for energy savings $r_n$ and the congestion status $c_{\mathcal{V}_n}$ of the channel which depends on the strategic choices of agent $n$ and its neighbors.

We want the reward for not forwarding $r_n$ to provide more benefit to a node that has low remaining energy. Hence nodes with a lower residual energy level are not inclined to broadcast while nodes with full batteries get a better utility for broadcasting.
Consequently, we define $r_n$ as the ratio of the energy already consumed $E_c$ to the initial available energy $E_0$: $r_n = E_c/E_0$.

\subsection{A game theoretical perspective}

In the game model corresponding to the utility-based policy of Eq.~\eqref{eq:payoff1}, the nodes are the players that choose among two strategies, {\it Forward} ($s_n=1$) and {\it Drop} ($s_n=0$). If the payoff of {\it Forward} is larger than the payoff of {\it Drop}, the sensor transmits, otherwise the sensor drops the packet. We consider that a sensor has always some data to send. The game is repeated for each packet transmission. Since $r_n$ is a function of the energy consumption at node $n$, the equilibrium of each repeated game is changing.
In this section we derive the mixed-strategy Nash-Equilibrium (NE) for one instance of the game, i.e. having $r_n$ and $\alpha_n$ values fixed. Since the game is a finite strategic form game, a mixed-strategy NE exists \cite{gametheory}.

Our utility-based policy is inspired by the Santa Fe Bar Problem (SFBP)\cite{mishra} where a congested resource, the bar, is shared by a set of agents, the bar customers. The customers enjoy their evening at the bar only if it is not over crowded (i.e. the capacity of the bar is lower than a fixed limit $c$).
The main difference with the forwarding game is that a player's decision impacts the congestion of the channel for only a subset of the players, i.e. the neighboring nodes. Moreover, due to the overlapping of coverage areas, some nodes contribute to several `channels'. Therefore, some nodes in a set of neighbor nodes sharing the channel may sense a congested channel while others see it free. As a consequence, the nodes that receive the same packet to forward do not necessarily have the same view of the level of congestion of the channel. Further, in the SFBP problem, $r_n=0$ and there is no reward for not attending the bar. We recall that this reward is introduced to account for the energy depletion of a node and reduce its incentive to forward and cooperate when its energy is low.

\subsubsection{Nash-Equilibrium analysis}
Given $\{\alpha_n$,$r_n\}_{n\in [1..N]}$, the expected payoff a sensor $n$ gets from selecting the action {\it Forward} is given by
\begin{equation}
 \mathbf{E}[u(1, c_{\mathcal{V}_n})]=\alpha_n.P(c_{\mathcal{V}_n}<c) + (\alpha_n-1).P(c_{\mathcal{V}_n}\geq c)
\end{equation}
with $P(c_{\mathcal{V}_n}<c)$ (resp. $P(c_{\mathcal{V}_n}\geq c)$) the probability that the channel is free (resp. the channel is congested).
Since $P(c_{\mathcal{V}_n}<c) = 1-P(c_{\mathcal{V}_n}\geq c)$, we have:
\begin{equation}
 \mathbf{E}[u(1, c_{\mathcal{V}_n})]=\alpha_n-P(c_{\mathcal{V}_n}\geq c)
\end{equation}

The expected payoff for playing {\it Drop} is given by:
\begin{equation}
\mathbf{E}[u(0, c_{\mathcal{V}_n})]=r_n
\end{equation}

A sensor will maximize its own payoff by forwarding if $ \mathbf{E}[u(1, c_{\mathcal{V}_n})]> \mathbf{E}[u(0, c_{\mathcal{V}_n})]$. Thus, we have:
\begin{equation}
\alpha_n-P(c_{\mathcal{V}_n}\geq c)>r_n
\end{equation}

We consider a mixed-strategy equilibrium where each sensor $n$ has a different equilibrium probability $p_n$ of playing {\it Forward}. This assumption is due to the fact that in a realistic network, the reward $r_n=E_c/E_0$ is likely to be different for every node as it is a function of the traffic a node has transmitted previously. The set of probabilities of forwarding $p_n$ can be expressed by:

\begin{equation}
p_n = {\rm Prob}\left[ \alpha_n-P(c_{\mathcal{V}_n}\geq c)>r_n \right]
\label{eq:pn}
\end{equation}

Since $r_n$ is fixed for all the nodes trying to access the channel at the same time, the equilibrium is completely determined by the values of the reward of forwarding $\alpha_n$ of all the nodes. If a clear relation between the vector of $\alpha_n$ and the $p_n$ can be determined, the values of the forwarding rewards can be chosen such as for instance to maximize the probabilities of forwarding of the nodes knowing the distribution of the energy rewards of the nodes.

As shown by Eq.\eqref{eq:pn}, $p_n$ is a function of $P(c_{\mathcal{V}_n}\geq c)$, the probability of the channel being congested. This congestion probability is a function of the forwarding probabilities of all the neighbor nodes $\mathcal{V}_n$ of $n$. For each neighbor node, its forwarding probability also depend on its own values of $\alpha_n$, $r_n$ and on the forwarding probabilities of its respective neighbor nodes. Hence, we have $p_n = f(\alpha, r)$, where $\alpha = [\alpha_1, ..., \alpha_n]$ and $r=[r_1, ..., r_N]$.
Further, the $p_n$ values are also strongly influenced by the medium access control (MAC) protocol which modifies $P(c_{\mathcal{V}_n}\geq c)$ by properly scheduling the transmissions.
Hence, deriving the distribution of the congestion probability is very complex since it depends on the distribution of the network, the physical transmission properties which determines the set of overlapping coverage areas, the MAC layer implementation and the flows being transmitted in the network.

Consequently, the Nash Equilibrium solution for the forwarding probabilities in \eqref{eq:pn} has little practical value, as not enough information is available to accurately characterized all the parameters. Moreover, since the game is repeated with different $r_n$ parameters, new probabilities have to be computed for each transmission. In the following, we propose a distributed heuristic approach where the value of $\alpha_n$ is updated by the node $n$ throughout its lifetime by measuring and interpreting the activity on the channel.

\subsection{The Distributed U-GRAB Heuristic}

\subsubsection{Congestion measure}
The exact number of concurrent transmissions on the wireless channel at time $t$ can not be determined by a sensor. The sensor's radio has only a partial view of the channel occupancy. However, compared to the problem where the players don't know the number of consumers that will attend the bar, we have a first valuable hint on the congestion level of the channel at the time of the decision. This hint is the `busy channel' information when listening to the channel just before sending a packet. Therefore we know that at least one sensor is already transmitting. When multiple channels are available (i.e for FDMA, CDMA systems), a channel per frequency/code can be considered. Therefore, a node can quantify how many channels are busy even though is does not know how many other sensors access each channel. Note however that the channel changes dynamically, and a channel sensed free, can still lead to collisions at the time of transmission. This is why we still have to account for the congestion probability in the equilibrium analysis provided previously.

In the proposed algorithm, we consider the sensed level of congestion $c_n$ as an estimate on the real level of congestion $c_{\mathcal{V}_n}$ of the network. We consider that a node chooses rationally its strategy as follows:
\begin{itemize}
\item if  $c_n  < c$, the network is considered as not congested and the payoffs for forwarding  (i.e. $u(1, c_n) = \alpha_n$) and not forwarding (i.e. $u(0, c_n) = r_n$) are computed,
\item if $c_n \geq c$, the network is considered as congested and the payoffs for forwarding (i.e. $u(1, c_n) =\alpha_n - 1)$ and not forwarding (i.e. $u(0, c_n) = r_n$) are computed.
\end{itemize}
The node chooses the strategy that maximizes its payoff knowing its estimate of $c_n$. Whenever $u(1,c_n)=u(0,c_n)$, the sensor flips a fair coin to decide whether it should forward or not.
When only one channel is considered, i.e. $c=1$, the above algorithm resumes for the case $c_n \geq c$ to directly choosing the strategy {\it Drop}. Consequently, payoff values do not have to be computed. In the following, we consider the case $c=1$.

\subsubsection{Choice of $\alpha_n$}
We recall that if the channel is not congested, a sensor transmits if and only if $\alpha_n > r_n$, i.e. when the reward for forwarding is higher than the energy savings reward. $\alpha_n$ is interpreted in this heuristic as an energy threshold that allows a sensor to forward a packet or not, depending on the amount of energy remaining in its battery.
When a sensor senses the channel to be free, the payoff function allows it to broadcast packets until $\alpha_n.100$ \% of its energy is consumed. The sensor stops broadcasting packets whenever the energy reward becomes higher than $\alpha_n$.

Once the energy level has reached $\alpha_n$ and the sensor has stopped broadcasting packets, it is allowed to increase its energy threshold $\alpha_n$ and resume broadcasting if it notices that its neighbor nodes do not forward any other messages he had received since he stopped forwarding. In this case, it believes that its neighbors with costs lower than its own cost do not forward anymore because their energy level is too low, too.

The value of the energy threshold of a sensor $n$ obtained after $k$ threshold increases, $\alpha_n(k)$, is computed according to $\alpha_n(k) = 1-x_0.q^{k} $ where $q \in [0,1]$ and $x_0 \in [0,1]$, providing a first energy threshold $\alpha_n(0) = 1-x_0$. As shown previously, the equilibrium of the problem depends on the value of $\alpha_n$ and $r_n$, an consequently on the choice of $x_0$ and $q$. The latter values are difficult to assess analytically. Therefore, we have chosen $q=0.75$ and $\alpha_n(0)=0.25$ empirically after several tests to provide the best possible energy statistics.

A sensor decides to increase $\alpha_n$ if it notices that no other neighbor with a lower cost forwards a packet. To detect such an event, each node counts the average number of packets $N_{high}$ received from its neighbor nodes with a packet cost $Q_P$ that is higher than its own cost $Q$ and the average number of packets $N_{low}$ received by its neighbor nodes with a lower packet cost $Q_P$. If $N_{high}=0$ there is no traffic on the network. But if $N_{high}>0$ and $N_{low}=0$, the current node gets packets to forward that its one hop neighbors do not forward.
The values for $N_{high}$ and $N_{low}$ are estimated at runtime using an exponential moving average.

Note that there is a particular transmission scenario for a homogeneous network that leads to an oscillatory behavior. Such behavior is encountered when all the nodes have the same level of energy (the reward is constant $r_n=r$), the same values of $\alpha_n$ and they all share the same channel. In this case, all the nodes sense a free channel and transmit concurrently. Thus, the channel gets congested and the sensors decide not to forward in the next transmission trial. As the channel becomes free again, the nodes resume forwarding and all the messages collide again. Such a behavior is neither fair nor efficient as a sensor never gets access to the channel. However, this scenario is unlikely to arise in a real network because of 3 main reasons. Firstly, as also considered in GRAB, the medium access protocol in this implementation follows a simple backoff procedure without acknowledgement. Hence, since the routing protocol is not slotted and a CSMA MAC is considered, the probability that the transmissions collide is reduced. Second, the fact that the nodes do not have the same view on the channel congestion due to overlapping of coverage areas further limits the occurrence of such a scenario. Thirdly, the condition $r_n=r$ is met when the network is newly started ($r_n=0$). But as the GRAB protocol starts with a first cost field setup stage, the distribution of the $r_n$ values is not uniform anymore as the data broadcasting stage is launched.

In terms of overhead, U-GRAB needs no additional information exchange compared to GRAB since a node overhears the packets exchanged on the channel to adjust its transmission parameters. The cost field setup bears the same cost than a regular gradient broadcasting algorithm. In the forwarding stage, no packet exchanges are needed to decide whether $\alpha_n$ has to be increased of not. In fact, a node listens to the packets on the channel and if it notices that no neighbor node forwards packets originating from nodes with higher costs, it increases $\alpha_n$.

\section{The UP-GRAB algorithm}\label{sec:PGGRAB}

The goal of the UP-GRAB algorithm is to account for both the channel congestion and the interference impact of the forwarding decision. In UP-GRAB, the interference avoidance property of P-GRAB is introduced in the collision aware model of U-GRAB.
UP-GRAB is based on almost the same utility policy as U-GRAB. The only difference is that a player $i$ chooses its strategies $s_i$ in the strategy set $S =\{0, 1\}$ where:
\begin{itemize}
\item $s_i=1$ corresponds to the action {\it Forward with probability $P_{IA}(i)$},
\item $s_i=0$ corresponds to the action {\it Not Forward}.
\end{itemize}

Hence, when the channel is considered as free, instead of simply broadcasting its packet, a node $n$ accounts for its interference potential and forwards a packet with probability $P_{IA}(i)$ as defined in Eq.\eqref{eq:fIAErfc}.
This probability of interference avoidance relies on the same neighborhood density measure $\Delta(i)$ and conversion function $f_{\rm erfc}$ as in P-GRAB.
A node is also able to adjust $P_{IA}(i)$ to the dynamics of the network: every time a congestion is detected, a node reduces $P_{IA}(i)$. Similarly, every time a free channel is detected, a node increases $P_{IA}(i)$. The increase or decrease of $P_{IA}$ is performed by modifying the value of the spreading factor $K$ of the conversion function $f_{\rm erfc}$. Hence, each node $i$ has its own value of the spreading factor denoted $K_i$.

As it can be seen on Fig.\ref{fig:fIAErfc}, when modifying $K_i$, a sensor can only choose a value of $P_{IA}$ that belongs to a bounded interval limited by the values of the functions $f_{\rm erfc}(\Delta(i))$ obtained for $K_i=1$ and for $K_i=+\infty$. For $K_i=1$, the function is close to a ramp function while for high values of $K_i$, $f_{\rm erfc}$ is close to a linear function of $\Delta(i)$.

Hence, if a congestion is detected, a node $i$ reduces $P_{IA}(i)$ by choosing an appropriate value for $K_i$ knowing its value of $\Delta(i)$. In this particular case, if the node has a smaller relative interference impact ($\Delta(i)<c$) it decreases $P_{IA}(i)$ by decrementing $K_i$. If the node has a higher relative interference impact ($\Delta(i)>c$), it has to increment $K_i$ to decrease $P_{IA}(i)$.
In the same way, upon detecting a free channel, a sensor increases $P_{IA}$ by decrementing $K_i$ (resp. incrementing $K_i$) if it has $\Delta(i)<c$ (resp. $\Delta(i)>c$).

To summarize, once a node $i$ has a packet to broadcast, it measures the channel state. Depending on its measure, it increments or decrements its spreading factor $K_i$ to either increase or decrease $P_{IA}(i)$. Then, it computes its payoff values according to Eq.\ref{eq:payoff1}. If the action {\it `Forward with probability $P_{IA}(i)$'} provides the highest payoff, it decides to broadcasts its packet with probability $P_{IA}(i)$. If the action {\it `Do not forward'} yields the best payoff, it does not transmit.
Note that the fair energy mechanism of the game in UP-GRAB is identical to U-GRAB since the payoff values of the game, the $\alpha_i$ and $r_i$ values, are calculated and updated according to the same algorithm.

\section{Simulation Results}\label{sec:results}

\subsection{Simulations setup}\label{subsec:simSetup}
For all the simulation results presented in this paper, a set of 100 different network configurations is considered where each network is composed of $N=1000$ randomly distributed nodes on an area of $500\times500$ meters.
Each sensor follows the specifications of a MICA2 Mote \cite{Mica2}. For each run, 30 randomly positioned events are created, triggering about 30 +/- 5 messages sent to the sink node.
Failures are modeled in a first approximation as a uniform probability where a node fails with probability $p_f$ when transmitting a packet, due to either fading or hardware malfunction. In this case, all the nodes are affected identically by the outage. Even if such uniform error distribution is not realistic, it provides a good first assessment of the performance of the algorithms.

The protocols have been implemented using the OMNet++ simulator in a modified version of the SENSIM sensor network simulator presented in \cite{LSUsimulator}. We enhanced the SENSIM simulator by adding a realistic radio layer model that accurately accounts for collisions originating from congestion and interference. In this layer, a node is able to completely demodulate a packet if and only if the Signal to Noise and Interference Ratio (SINR) is high enough during the whole packet reception duration. We also developed and use here a basic CSMA MAC layer where a node sends its packet after a random waiting time. There is no feedback on local network congestion at the MAC level. The overhead required by each protocol is included in the results presented hereafter.

The transmission robustness for one run is measured by the message success ratio, i.e. the ratio of the number of correctly received messages to the number of sent messages. The transmission delay is the average of the minimum delay needed by all the successfully received messages to reach the sink (given in milliseconds). The energy consumption is measured herein using two metrics. First, we consider the percentage of the initial amount of energy consumed by the nodes at the end of the simulation. Secondly, we measure the number of forwarded packets by all the nodes during the run. Since 100 random network configurations are tested, average and standard deviation values are reported for all the metrics.

\begin{figure}
  \includegraphics[width=3.5in]{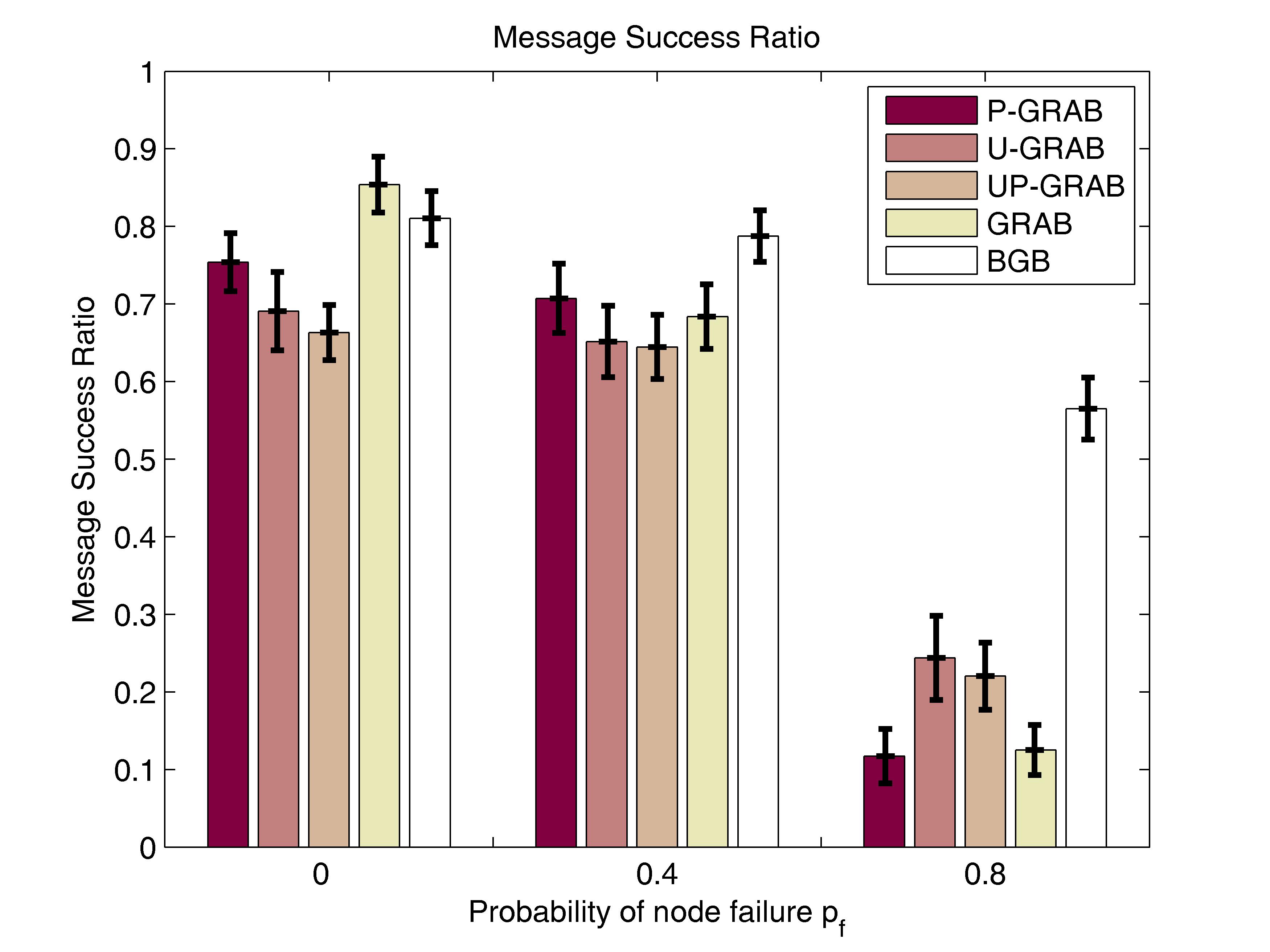}
  \vspace{0.1in}
  \includegraphics[width=3.5in]{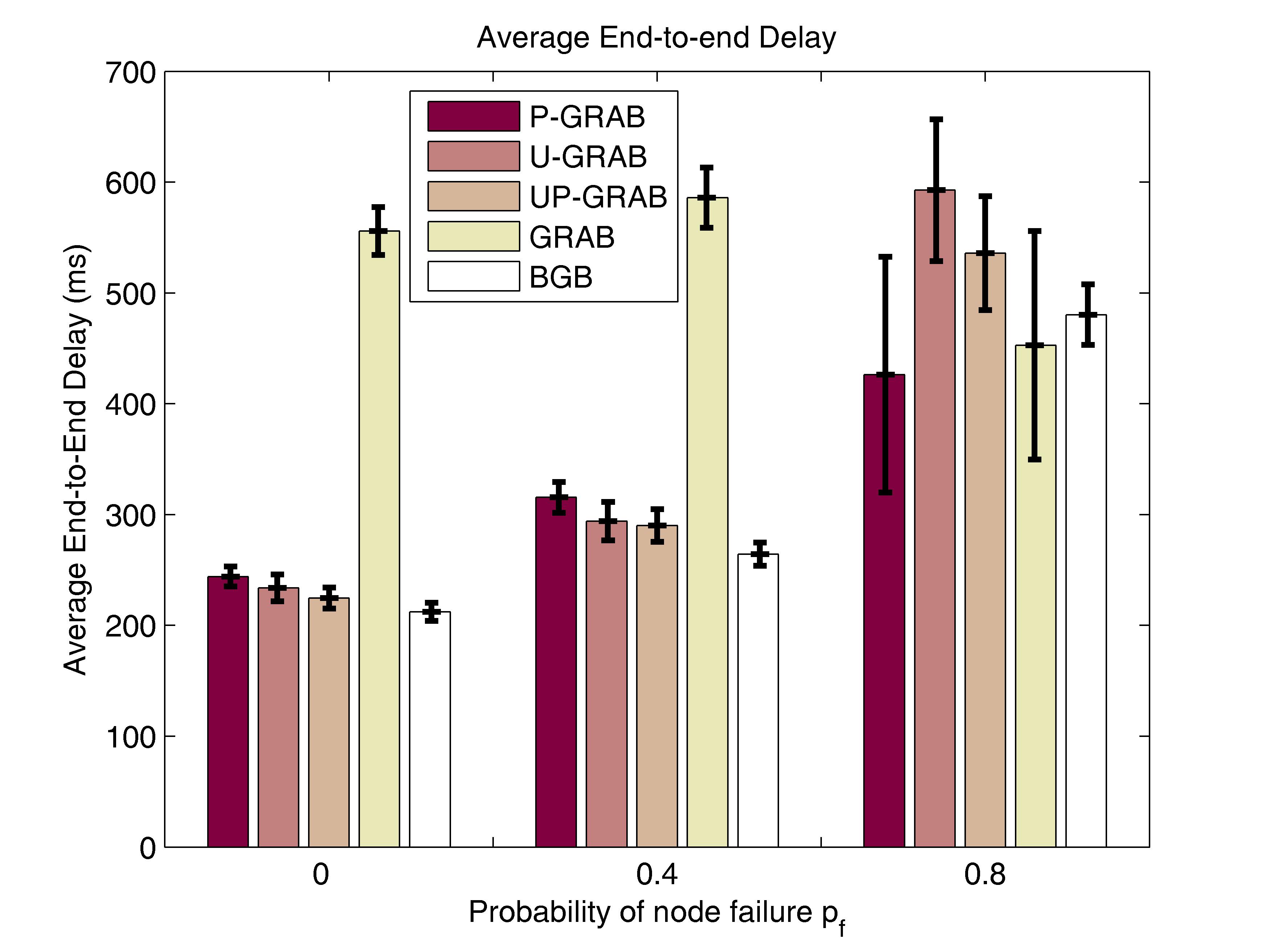}
    \caption{Performance of P-GRAB, U-GRAB, UP-GRAB, GRAB ($F_\alpha = 10$) and BGB for $p_f = \{0, 0.4, 0.8\}$ in terms of robustness (top) and end-to-end delay (bottom).}
    \label{fig:resGGRAB-RobDel}
\end{figure}

\begin{figure}
  \includegraphics[width=3.5in]{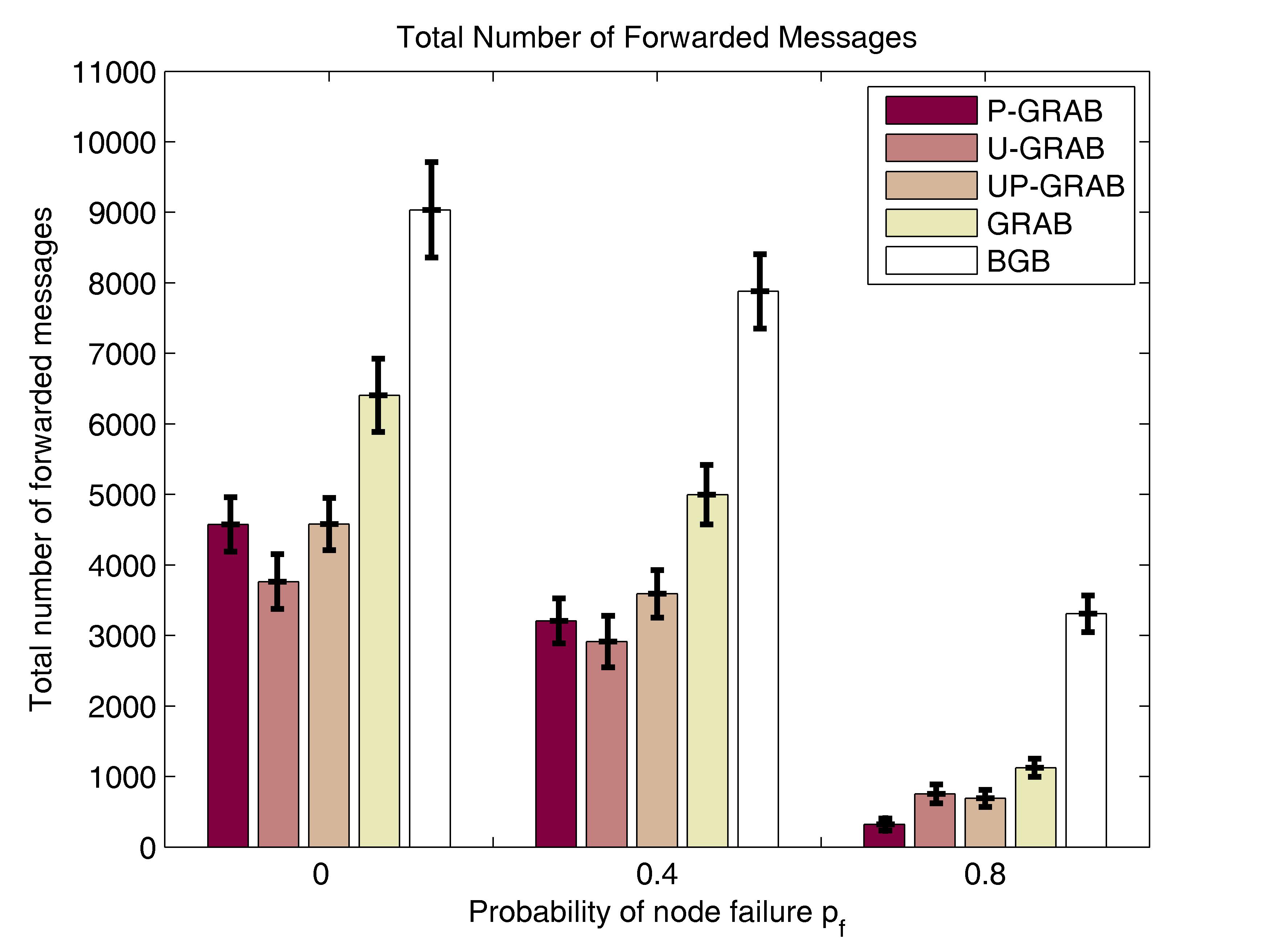}
  \vspace{0.1in}
  \includegraphics[width=3.5in]{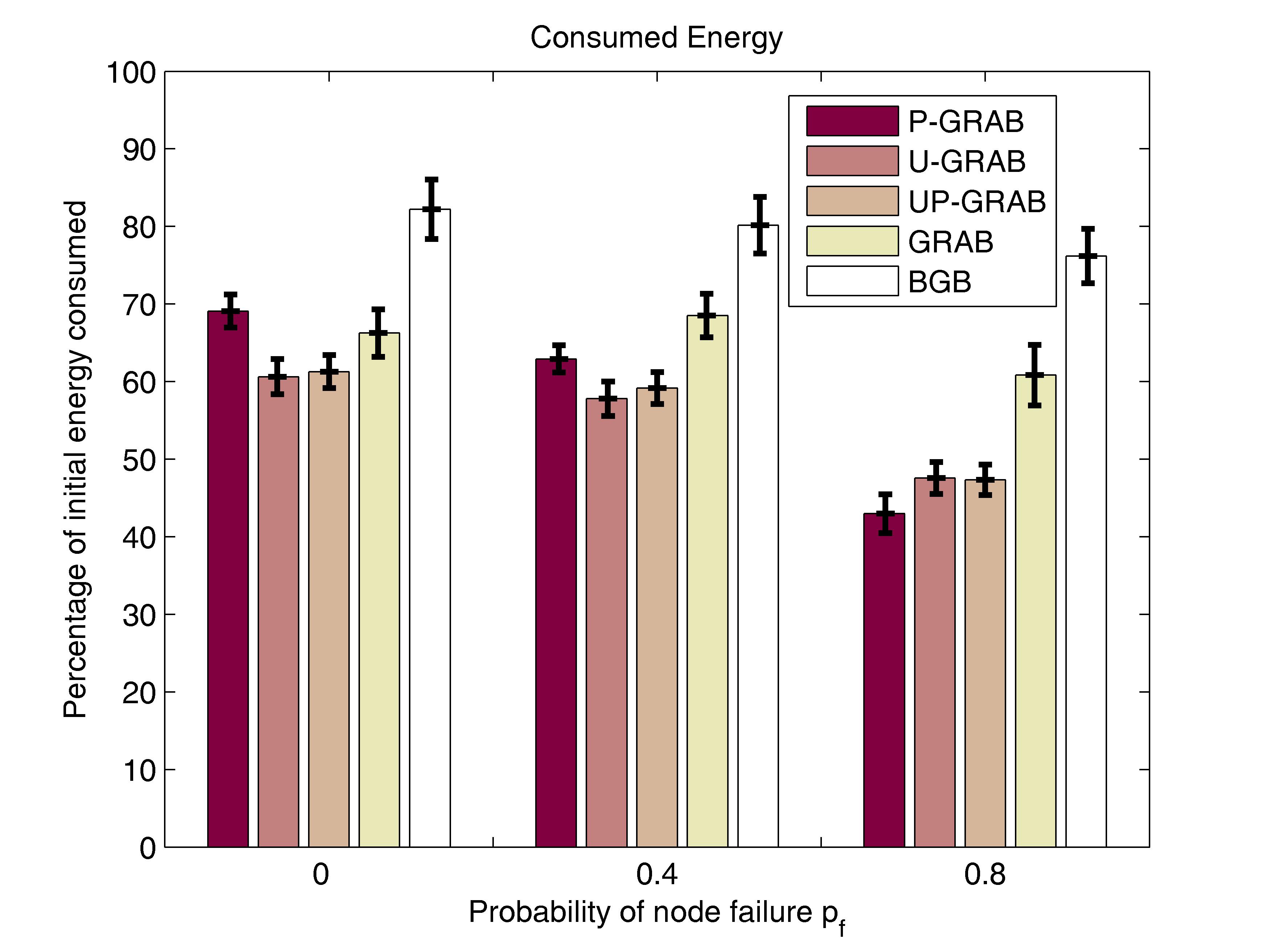}
    \caption{Performance of P-GRAB, U-GRAB, UP-GRAB, GRAB ($F_\alpha = 10$) and BGB for $p_f = \{0, 0.4, 0.8\}$ in terms of the forwarding load (top) and the average energy consumption (bottom).}
    \label{fig:resGGRAB-ForEn}
\end{figure}

\begin{figure}
  \includegraphics[width=3.5in]{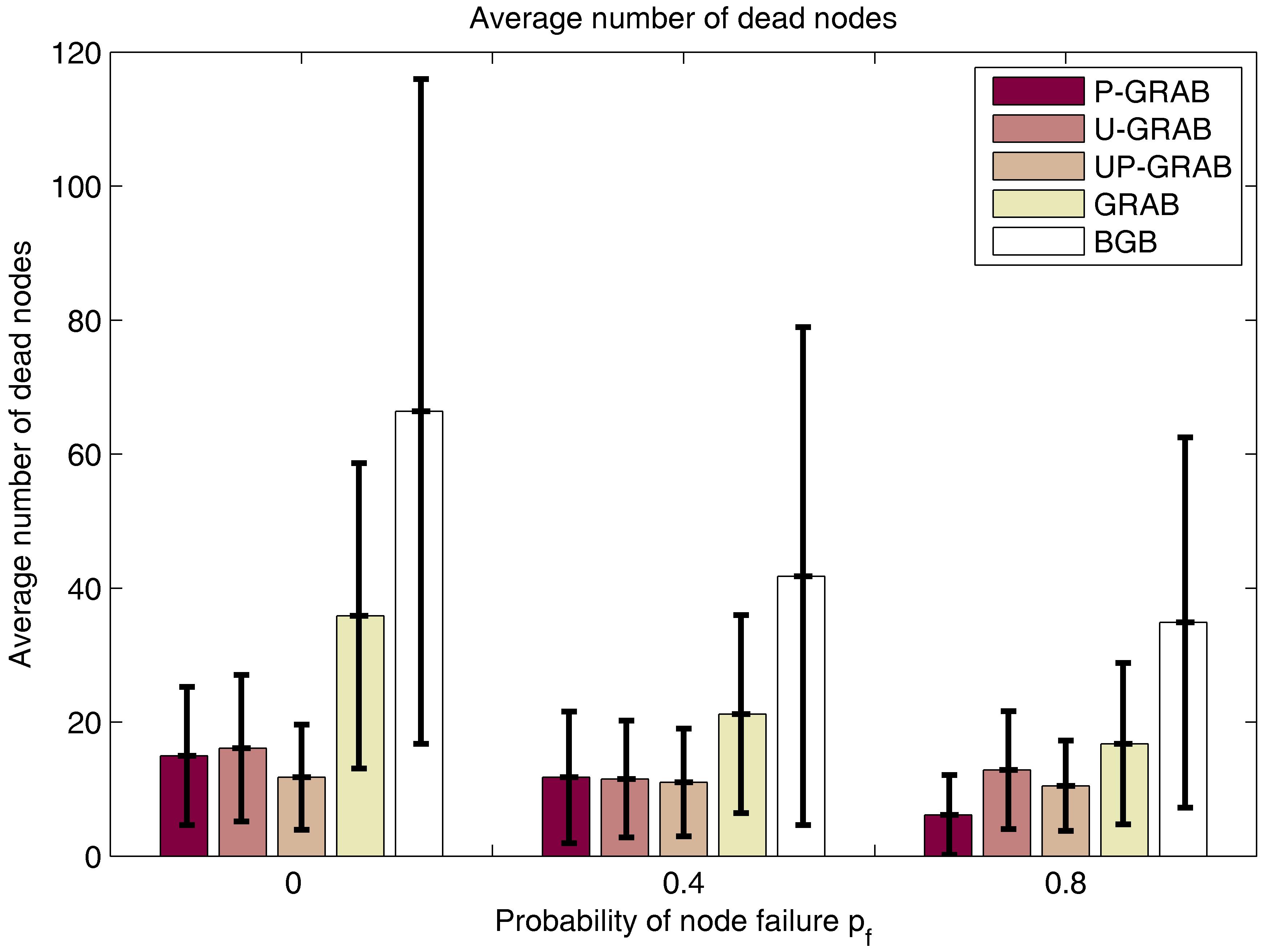}
        \caption{Performance of P-GRAB, U-GRAB, UP-GRAB, GRAB ($F_\alpha = 10$) and BGB for $p_f = \{0, 0.4, 0.8\}$ in terms of the average number of dead nodes.}
    \label{fig:resGGRAB-Dead}
\end{figure}

In Fig.~\ref{fig:resGGRAB-RobDel},~\ref{fig:resGGRAB-ForEn} and \ref{fig:resGGRAB-Dead}, we present the results obtained for P-GRAB, U-GRAB, UP-GRAB, GRAB and BGB. A credit factor of $F_\alpha = 10$ is considered for GRAB which provides the best compromise between robustness and delay.
According to Section \ref{subsec:Kvalue-PGRAB}, a uniform value of $K=2$ is set for P-GRAB. Figure \ref{fig:resGGRAB-RobDel} shows robustness and delay performance metrics. Figure \ref{fig:resGGRAB-ForEn} provides energy consumption metrics and Fig.~\ref{fig:resGGRAB-Dead} gives the average number of nodes that are dead at the end of the simulation.
In this section, we analyze first the performance of our three broadcasting strategies and then compare them to GRAB and BGB.

\subsection{Performance of P-GRAB, U-GRAB and UP-GRAB}

\paragraph{\rm For $p_f\leq0.4$}
In terms of robustness, P-GRAB has a message success ratio that is about 15\% higher than U-GRAB or UP-GRAB.
With respect to the number of forwarded messages in Fig.\ref{fig:resGGRAB-ForEn}, UP-GRAB and P-GRAB use about 24\% more packets than U-GRAB to transmit the same amount of data. Since P-GRAB has a higher message success ratio than UP-GRAB, we can state that P-GRAB takes better forwarding decisions than UP-GRAB by reducing the occurrence of collisions. For U-GRAB, its reduced robustness may not completely originate from a higher number of collisions, but rather be a consequence of a smaller overall number of transmitted packets. However, it is clear that the probabilistic strategy better accounts for the interference than the utility-based strategies.

Energy consumption statistics are also given in Fig.\ref{fig:resGGRAB-ForEn}. The energy model in the simulations accounts for both the energy consumed for reception and transmission. Therefore, the average consumed energy is not a direct mapping of the number of forwarded messages. Although UP-GRAB sends more packets than U-GRAB, it results in the same average energy consumption. It is consistent with the fact that more packets are lost by UP-GRAB and there is no energy consumed for their reception. For P-GRAB, the increase in the number of forwarded packets is clearly shown in the consumed energy figure where an increased energy expenditure is observed.
Regarding delay, the three strategies provide the same average transmission delay which is of about the same order as the delay of BGB. BGB provides the shortest delay as all the available paths are available in its implementation.
The three algorithms also have the same average number of dead nodes which assesses that they have a similar impact on the distribution of the energy consumption among the nodes of the network.

\paragraph{\rm For $p_f=0.8$}
For unreliable networks, the utility-based approaches perform much better than the probabilistic approach. Two times more messages arrive with U-GRAB and UP-GRAB compared to P-GRAB. When $p_f=0.8$, the geometry of the network is very different from the one used to compute the values of $\Delta(i)$ as nodes fail 80\% of the time. Furthermore, the neighborhood distribution changes very frequently with the nodes getting on and off because of outages. Therefore, the interference impact of a transmission is overestimated and too few packets are broadcasted by P-GRAB.
The average delay of the transmission is greatly increased in such unreliable network. With nodes frequently failing, paths using shorter hops are more successful as they see more nodes contributing to the transmission and are therefore less sensible to the elevated outage. The delays are longer for U-GRAB and UP-GRAB which is consistent with the increased message success ratio they provide. These utility-based approaches provide longer but more reliable transmissions.
The increased robustness of U-GRAB and UP-GRAB impacts their energy consumption as more broadcasts are performed. However, the increase in energy consumption is small compared to the 100\% gain in message success ratio they offer.

As a conclusion, we can state that so far, the interference mitigation capability of UP-GRAB adds little contribution to the performance of U-GRAB. It is the congestion mechanism common to U-GRAB and UP-GRAB that mostly impacts the broadcasting decision of UP-GRAB, hence triggering similar outcome.
We can also stress that P-GRAB performs very well for networks with low to middle probability of failures. When the occurrence of failures is very high, U-GRAB provides a much better outcome than P-GRAB as it better adjusts to the frequent local changes of the network distribution. Furthermore, for a stable network, P-GRAB is the best option to efficiently transmit data quickly and with low energy consumption. Once the network gets unreliable as it happens towards the end of its life, U-GRAB becomes more efficient since its dynamic properties have a much better outcome on robustness and energy consumption.

\subsection{Comparison with GRAB and BGB}

Based on the conclusions we draw in the previous section, we compare GRAB to P-GRAB for $p_f\leq0.4$ and GRAB to U-GRAB for $p_f=0.8$.

\paragraph{\rm For $p_f\leq0.4$}
The robustness of GRAB still exceeds the robustness of P-GRAB when $p_f=0$. However, when the failure rate increases to $p_f=0.4$, P-GRAB outperforms GRAB in robustness. Furthermore, P-GRAB needs up to 30\% less packets transmissions and up to 18\% less energy to transmit its data, which results in a far better robustness/energy tradeoff.
The energy consumption is also better spread among the nodes as 2 times less nodes die at the end of the simulations with P-GRAB.
It is in terms of delay that P-GRAB greatly outperforms GRAB, needing about half the time to transmit the data in average, which is really significant and close to the delay of BGB. BGB provides here the best robustness and delay, but clearly to the price of a higher energy cost (BGB uses up to 100\% more forwarded messages and needs 26\% more energy than P-GRAB).

\paragraph{\rm For $p_f=0.8$}
U-GRAB really outperforms GRAB in such unreliable environments as its message success ratio is two times higher than the one of GRAB. It also needs 30\% less transmissions and 21\% less energy to achieve such a good robustness. The drawback is an increased average delay as U-GRAB spends 25\% more time to transmit its data. However, for these severe transmission conditions, the increase in robustness is very beneficial which makes higher transmission delays acceptable. Also, compared to BGB, U-GRAB offers a decent robustness, a similar delay but needs 75\% less packets, resulting in 37\% in energy savings.

\section{Conclusions}

In this paper we have proposed three different solutions to improve the performance of gradient broadcasting for sensor networks through the use of an interference impact measure and a congestion-aware utility policy. Our aim was to improve the robustness, energy consumption and transmission delay for networks deployed in harsh environments, prone to a high probability of node and communication failures.
We have observed via simulations that all the proposed strategies greatly outperform GRAB in terms of the average transmission delay, reducing it by up to two times.
The probabilistic approach, P-GRAB, which accounts for the interference impact of a node transmission in the broadcasting decision, experiences the best performance for high to average reliability of the nodes. For such networks, the knowledge of the topology greatly improves the robustness-energy consumption trade-off compared to GRAB. For unreliable networks, where topology varies locally because of increased number of failures, the utility based GRAB (U-GRAB) is the best algorithm, as it adapts to the instantaneous congestion of the network, providing a better collision avoidance scheme than GRAB or P-GRAB.
In this case, cooperation clearly improves the protocol's reactivity.

Since the combined interference and congestion aware protocol UP-GRAB does meet our expectations, we will concentrate in the future in providing a general broadcasting framework where U-GRAB and P-GRAB are used alternatively or concurrently in different parts of the network depending on the network stability. Hence, the sink becomes a major player which decides which algorithm to use depending on its view of the network performance (occurrence of failures, packet drop rate, etc...).

\section*{Acknowledgments}
This work was supported in part by the Marie Curie OIF Action of the European Community's Sixth Framework Program (DistMO4WNet project) and by the ONR grant \#N00014-06-1-0063. This article only reflects the author's views and neither the Community nor the ONR are liable for any use that may be made of the information contained herein. This work was
presented in part to IEEE ISWPC, Santorini, Greece in May 2008 and to IEEE Milcom, Boston, MA, USA in October 2009.

\bibliographystyle{IEEEtran}
\bibliography{References}

\end{document}